\documentstyle[12pt,prd,aps,epsfig,preprint]{revtex}
\begin{document}
\draft \tightenlines
\date{\today}
\title{Scalar mesons in radiative  $\phi\rightarrow K^0\overline{K}^0\gamma$ decay}

\author{A. Gokalp~\thanks{agokalp@metu.edu.tr},  C. S. Korkmaz and O. Yilmaz~\thanks{oyilmaz@metu.edu.tr}}
\address{ {\it Physics Department, Middle East Technical University,
06531 Ankara, Turkey}} \maketitle

\begin{abstract}
We study the radiative $\phi\rightarrow K^0\overline{K}^0\gamma$
decay within a phenomenological framework by considering the
contributions of the $f_{0}(980)$ and $a_{0}(980)$ scalar
resonances. We calculate the branching ratio $B(\phi\rightarrow
K^0\overline{K}^0\gamma)$ by employing the coupling constants
$g_{f_0K^{+}K^{-}}$ and $g_{a_0K^{+}K^{-}}$ as determined by
different experimental groups.
\end{abstract}

\thispagestyle{empty} ~~~~\\ \pacs{PACS numbers: 12.20.Ds, 13.20.Jf,
13.40.Hq}
\newpage
\setcounter{page}{1}
\section{Introduction}

The radiative decays of $\phi$ meson into a single photon and a pair
of neutral pseudoscalar mesons are valuable sources of information
in hadron physics. These days can provide insight into the structure
and the properties of low-mass scalar resonances. In particular, the
radiative decay $\phi\rightarrow K^0\overline{K}^0\gamma$ where the
scalar resonances $f_{0}(980)$ and $a_{0}(980)$ are known to provide
the dominant contribution to the amplitude allows for a direct
measurement of the couplings of the $K\overline{K}$ system to
$f_{0}$ and $a_{0}$ scalar resonances and thus it can yield
information on the properties of these scalar resonances. Moreover,
the study of the radiative decay process $\phi\rightarrow
K^0\overline{K}^0\gamma$ is important because it provides a
background to the reaction $\phi\rightarrow K^0\overline{K}^0$. This
latter process was proposed as a way to study CP violation and
measuring the ratio $\epsilon^\prime/\epsilon$ \cite{R1}. Since this
involves seeking for very small effects, if the branching ratio of
the decay $\phi\rightarrow K^0\overline{K}^0\gamma$ is of the order
of $10^{-6}$, or more precisely if $B(\phi\rightarrow
K^0\overline{K}^0\gamma)\geq 10^{-6}$, then this background decay
$\phi\rightarrow K^0\overline{K}^0\gamma$ will limit the scope of CP
violation measurements in $\phi\rightarrow K^0\overline{K}^0$ decay
at DA$\Phi$NE. Therefore, the study of the reaction $\phi\rightarrow
K^0\overline{K}^0\gamma$ and the calculation of the branching ratio
$B(\phi\rightarrow K^0\overline{K}^0\gamma)$ is crutial for the
measurements of CP violation and small CP violating parameters in
$\phi\rightarrow K^0\overline{K}^0$ decay.

The decay $\phi\rightarrow K^0\overline{K}^0\gamma$ was first
considered by Achasov et al. \cite{R2} using the one-loop mechanism
where the decay proceeds through the chain of reactions as
$\phi\rightarrow K^+K^-\rightarrow (f_0+a_0)\gamma\rightarrow
K^0\overline{K}^0\gamma$. They noted the negative interference
between the contributions of $f_0$ and $a_0$ resonances, and they
obtained the value $BR(\phi\rightarrow (f_0+a_0)\gamma\rightarrow
K^0\overline{K}^0\gamma)=1.3\times 10^{-8}$ for the branching ratio
of the $\phi\rightarrow K^0\overline{K}^0\gamma$ decay for some set
of $f_0$ and $a_0$ masses and the values of the coupling constants
$g_{f_0K^+K^-}$ and $g_{a_0K^+K^-}$.

The radiative $\phi$-meson decays, among other radiative decay
processes of the type $V^0\rightarrow P^0P^0\gamma$, where V and P
belong to the lowest multiplets of vector (V) and pseudoscalar (P)
mesons, were studied by Bramon et al. \cite{R3} in the framework of
vector meson dominance (VMD) mechanism using standard Lagrangians
obeying SU(3)-symmetry. In this framework, the radiative
$\phi\rightarrow K^0\overline{K}^0\gamma$ reaction proceeds through
the decay chains $\phi\rightarrow K^0V\rightarrow
K^0\overline{K}^0\gamma$ where the intermediate vector mesons are
$V={K^*}^0$ and $V=\overline{K^*}^0$. They obtained the branching
ratio for the decay as $B(\phi\rightarrow
K^0\overline{K}^0\gamma)=2.7\times 10^{-12}$. Bramon et al.
\cite{R4} later considered the radiative vector meson decays in
Chiral Perturbation Theory (ChPT) using chiral effective Lagrangians
enlarged to include on-shell vector mesons, and they calculated the
branching ratios for various decays of type $V^0\rightarrow
P^0P^0\gamma$ at the one-loop level, including both $\pi\pi$ and
$K\overline{K}$ intermediate loops. In this approach the decay
$\phi\rightarrow K^0\overline{K}^0\gamma$ proceeds through
charged-kaon loops, and they obtained the contribution of
charged-kaon loops to the branching ratio as $B(\phi\rightarrow
K^0\overline{K}^0\gamma)=7.6\times 10^{-9}$. They concluded that due
to smallness of the VMD contribution, this decay is dominated by
charged-kaon loops.

An additional contribution to the decay of the type $V^0\rightarrow
P^0P^0\gamma$ is provided by the amplitude involving scalar mesons
as an intermediate state. These radiative decays were studied by
Marko et al. \cite{R5} in the framework of unitarized chiral
perturbation theory. They used the techniques of chiral unitary
theory developed earlier to include the final state interactions of
two pseudoscalars by summing the pseudoscalar loops through the
Bethe-Salpeter equation. A review is given by Oller et al.
\cite{R6}. In this approach the scalar resonances are generated
dynamically by utilizing the one-loop pseudoscalar amplitudes. The
reaction $\phi\rightarrow K^0\overline{K}^0\gamma$ was studied by
Oller \cite{R7} within this framework. The amplitude of this
radiative decay has contributions coming from the scalar mesons
$f_0(980)$ and $a_0(980)$, and Oller obtained for the branching
ratio the value $B(\phi\rightarrow K^0\overline{K}^0\gamma)=5\times
10^{-8}$. The radiative vector meson decays $V^0\rightarrow
P^0P^0\gamma$ were also studied using Linear Sigma Model
(L$\sigma$M), which is a $U(3)\times U(3)$ chiral model that
incorporates the pseudoscalar and scalar meson nonets \cite{R8}. The
decay $\phi\rightarrow K^0\overline{K}^0\gamma$ was studied within
the framework of L$\sigma$M by Escribano \cite{R9} who obtained the
result $B(\phi\rightarrow K^0\overline{K}^0\gamma)=7.5\times
10^{-8}$.

In this work, we attempt to calculate the branching ratio
$B(\phi\rightarrow K^0\overline{K}^0\gamma)$ using a
phenomenological approach employed earlier in the studies of
radiative $\phi\rightarrow \pi^0\pi^0\gamma$ \cite{R10} and
$\phi\rightarrow \pi^0\eta\gamma$ \cite{R11} decays. In our
calculation we use the kaon-loop model \cite{R2} where the initial
vector meson $\phi$ decays into a pair of charged kaons, which after
the emission of a photon couple to the neutral kaon pair $
K^0\overline{K}^0$ through the scalar resonances $a_0$ and $f_0$.
Moreover, we consider the vertices involving the scalar mesons as
point like, therefore the effects of the structure are reflected in
the coupling constants.

\section{Formalism}

The mechanism of the radiative decay process $\phi\rightarrow
K^0\overline{K}^0\gamma$ in the kaon-loop model is provided by the
reactions $\phi\rightarrow K^+K^-\gamma\rightarrow
K^0\overline{K}^0\gamma$ where the last reaction proceeds by a
two-step mechanism with the charged-kaon loop coupling to the final
$K^0\overline{K}^0$ state  through the scalar resonance $f_0(980)$
or $a_0(980)$. In Fig. 1 we show the processes contributing to the
$\phi\rightarrow K^0\overline{K}^0\gamma$ amplitude diagramatically
where the diagram in Fig. 1(c) results from the minimal coupling for
gauge invariance. We do not make any assumption about the structure
of the scalar mesons $S=f_0$ or $a_0$. We note that  the $\phi$
meson and the scalar mesons $f_0$ and $a_0$ both couple strongly to
the $ K^+K^-$ system, we therefore describe the $\phi K^+K^-$ and $S
K^+K^-$ vertices phenomenologically by effective Lagrangians. The
$\phi K^+K^-$ vertex in the diagrams shown in Fig. 1 is described by
the phenomenological Lagrangian

\begin{eqnarray}\label{e1}
{\cal L}_{\phi K^+K^-}=-ig_{\phi K^+K^-} \phi^\mu (
   K^{+}\partial_{\mu}K^{-}-K^{-}\partial_{\mu}K^{+})~~.
\end{eqnarray}
We utilize the experimental value for the branching ratio
$B(\phi\rightarrow  K^+K^-)$ \cite{R12} and determine the coupling
constant $g_{\phi K^{+}K^{-}}$ as $g_{\phi K^{+}K^{-}}=(4.47\pm
0.05)$. The $S K^{+}K^{-}$ vertex, where S denotes the scalar meson
$f_0$ or $a_0$, is described by the phenomenological Lagrangian
\begin{eqnarray}\label{e2}
{\cal L}_{S K^{+}K^{-}}= -g_{SK^+K^-}K^+K^-S~~,
\end{eqnarray}
which is usually considered to define the coupling constant
$g_{SK^+K^-}$ \cite{R13}. Similarly, we describe the
$SK^0\overline{K}^0$ vertex by the effective Lagrangian
\begin{eqnarray}\label{e3}
{\cal L}_{SK^0\overline{K}^0}=
-g_{SK^0\overline{K}^0}K^0\overline{K}^0S~~.
\end{eqnarray}
Furthermore, isotopic spin invariance implies that the coupling
constants $g_{S K^{+}K^{-}}$ and $g_{S K^{0}\overline{K}^{0}}$ are
related by the equations
$g_{f_0K^{+}K^{-}}=g_{f_0K^{0}\overline{K}^{0}}$ and
$g_{a_0K^{+}K^{-}}=-g_{a_0K^{0}\overline{K}^{0}}$ \cite{R2}. In our
approach we assume isotopic spin invariance and use the coupling
constants $g_{S K^{+}K^{-}}$ and $g_{S K^{0}\overline{K}^{0}}$
satisfying this requirement.

These coupling constants have been determined by theoretical
calculations and from experimental analysis. In our phenomenological
calculation we use the values of these coupling constants which are
determined from the experimental studies of the radiative decay
processes $\phi\rightarrow  \pi^0\pi^0\gamma$ and $\phi\rightarrow
\pi^0\eta\gamma$.

We then obtain the amplitude for the radiative decay reaction
$\phi\rightarrow K^0\overline{K}^0\gamma$  following from the
diagrams shown in Fig. 1 as
\begin{eqnarray}\label{e4}
{\cal M}(\phi\rightarrow K^0\overline{K}^0\gamma)&=&-\frac{e~g_{\phi
KK}}{i2\pi^{2}M_{K^+}^{2}}\left[(p\cdot k)(\epsilon\cdot
u)-(p\cdot\epsilon)(k\cdot u)\right] I(a,b) \nonumber \\
&&~~\times {\cal M}(K^+K^-\rightarrow K^0\overline{K}^0)
\end{eqnarray}
where $(u,p)$ and $(\epsilon, k)$ are the polarizations and
four-momenta of the $\phi$ meson and the photon, respectively. The
invariant function I(a,b) has been calculated in different contexts
\cite{R13,R14}, and it is given by
\begin{eqnarray}\label{eq5}
I(a,b)=\frac{1}{2(a-b)} -\frac{2}{(a-b)^{2}}\left [
f\left(\frac{1}{b}\right)-f\left(\frac{1}{a}\right)\right ]
+\frac{a}{(a-b)^{2}}\left [
g\left(\frac{1}{b}\right)-g\left(\frac{1}{a}\right)\right ]\label{6}
\end{eqnarray}

\begin{eqnarray}\label{eq6}
&&f(x)=\left \{ \begin{array}{rr}
           -\left [ \arcsin (\frac{1}{2\sqrt{x}})\right ]^{2}~,& ~~x>\frac{1}{4} \\
\frac{1}{4}\left [ \ln (\frac{\eta_{+}}{\eta_{-}})-i\pi\right
]^{2}~, & ~~x<\frac{1}{4}
            \end{array} \right.
\nonumber \\ && \nonumber \\ &&g(x)=\left \{ \begin{array}{rr}
        (4x-1)^{\frac{1}{2}} \arcsin(\frac{1}{2\sqrt{x}})~, & ~~ x>\frac{1}{4} \\
 \frac{1}{2}(1-4x)^{\frac{1}{2}}\left [\ln (\frac{\eta_{+}}{\eta_{-}})-i\pi \right ]~, & ~~ x<\frac{1}{4}
            \end{array} \right.
\nonumber \\ && \nonumber \\ &&\eta_{\pm}=\frac{1}{2x}\left [
1\pm(1-4x)^{\frac{1}{2}}\right ] ~,
\end{eqnarray}
where $a=M_{\phi}^{2}/M_{K^+}^{2}$ and $b=M_{KK}^{2}/M_{K^+}^{2}$
with $M_{KK}^{2}$ being the invariant mass of the final
$K^0\overline{K}^0$ system given by
$M_{K^0\overline{K}^0}^2=(p-k)^2=q^2$. The amplitude ${\cal
M}(K^+K^-\rightarrow K^0\overline{K}^0)$ contains the scalar $f_0$
and $a_0$ resonances and  in the approach we adopted it is given by
\begin{eqnarray}\label{e7}
{\cal M} (K^+K^-\rightarrow
K^0\overline{K}^0)=-ig_{SK^+K^-}g_{SK^0\overline{K}^0}\frac{1}{q^2-M_S^2}~~.
\end{eqnarray}

Since the scalar resonances $f_0$ and $a_0$ are unstable and they
have a finite lifetime we use Breit-Wigner propagators with an
energy dependent width for these resonances. We therefore in the
scalar meson propagator make the replacement $q^2-M_S^2\rightarrow
q^2-M_S^2+i\sqrt{q^2}\Gamma_S$  where
\begin{eqnarray}\label{e8}
  \Gamma_{f_0}(q^2)&=&\frac{g^2_{f_0K^+K^-}}{16\pi\sqrt{q^2}}\sqrt{1-\frac{4M_{K^+}^2}{q^2}}~\theta(\sqrt{q^2}-2M_{K^+})
  \nonumber \\
&+&\frac{g^2_{a_0K^0\overline{K}^0}}{16\pi\sqrt{q^2}}\sqrt{1-\frac{4M_{K^0}^2}{q^2}}~\theta(\sqrt{q^2}-2M_{K^0})
\nonumber \\
&+&\frac{2}{3}\Gamma_{f_0}\frac{M_{f_0}}{\sqrt{q^2}}
\frac{\sqrt{1-\frac{4M_{\pi^0}^2}{q^2}}}{\sqrt{1-\frac{4M_{\pi^0}^2}{M_{f_0}^2}}}~
\theta(\sqrt{q^2}-2M_{\pi^0})~~,
\end{eqnarray}
and
\begin{eqnarray}\label{e9}
  \Gamma_{a_0}(q^2)&=&\frac{g^2_{a_0K^+K^-}}{16\pi\sqrt{q^2}}\sqrt{1-\frac{4M_{K^+}^2}{q^2}}~\theta(\sqrt{q^2}-2M_{K^+})
  \nonumber \\
&+&\frac{g^2_{a_0K^0\overline{K}^0}}{16\pi\sqrt{q^2}}\sqrt{1-\frac{4M_{K^0}^2}{q^2}}~\theta(\sqrt{q^2}-2M_{K^0})
\nonumber \\
&+&\Gamma_{a_0}\frac{M_{a_0}}{\sqrt{q^2}}
\frac{\sqrt{\left[1-\frac{(M_{\pi^0}+M_\eta)^2}{q^2}\right]\left[1-\frac{(M_{\pi^0}-M_\eta)^2}{q^2}\right]}}
{\sqrt{\left[1-\frac{(M_{\pi^0}+M_\eta)^2}{M_{a_0}^2}\right]\left[1-\frac{(M_{\pi^0}-M_\eta)^2}{M_{a_0}^2}\right]}}~
\theta(\sqrt{q^2}-(M_{\pi^0}+M_\eta))~~,
\end{eqnarray}
and we use the experimental values for the widths $\Gamma_{f_0}$ and
$\Gamma_{a_0}$ \cite{R12} in the above expressions. Then the
differential decay probability for the radiative decay
$\phi\rightarrow K^0\overline{K}^0\gamma$ for an unpolarized $\phi$
meson at rest is given as
\begin{eqnarray}\label{e10}
\frac{d\Gamma}{dE_{\gamma}dE_{1}}=\frac{1}{(2\pi)^{3}}~\frac{1}{8M_{\phi}}~
\mid {\cal M}\mid^{2} ,
\end{eqnarray}
where E$_{\gamma}$ and E$_{1}$ are the photon and $K^0$ meson
energies respectively. We perform an average over the spin states of
$\phi$ meson and a sum over the polarization states of the photon.
The decay width $\phi\rightarrow K^0\overline{K}^0\gamma$ is then
obtained by integration
\begin{eqnarray}\label{e11}
\Gamma=\int_{E_{\gamma,min.}}^{E_{\gamma,max.}}dE_{\gamma}
       \int_{E_{1,min.}}^{E_{1,max.}}dE_{1}\frac{d\Gamma}{dE_{\gamma}dE_{1}}~~,
\end{eqnarray}
where the minimum photon energy is E$_{\gamma, min.}=0$ and the
maximum photon energy is given as
$E_{\gamma,max.}=(M_{\phi}^{2}-4M_{K_0}^2)/2M_{\phi}$. The maximum
and minimum values for the energy E$_{1}$ of $K^0$ meson are given
by
\begin{eqnarray}\label{e12}
\frac{1}{2(2E_{\gamma}M_{\phi}-M_{\phi}^{2})} \left\{
-2E_{\gamma}^{2}M_{\phi}+3E_{\gamma}M_{\phi}^{2}-M_{\phi}^{3}
 ~~~~~~~~~~~~~~~~~~~~~~~~~~~~\right. \nonumber \\
\pm  E_{\gamma}\sqrt{(-2E_{\gamma}M_{\phi}+M_{\phi}^{2})
       (-2E_{\gamma}M_{\phi}+M_{\phi}^{2}-4M_{K_0}^{2})}~\left\}\right.
       ~.\nonumber
\end{eqnarray}

\section{Results and Discussion}

The branching ratios $BR(\phi\rightarrow f_0(980)\gamma)$ and
$BR(\phi\rightarrow a_0(980)\gamma)$ are listed as
$BR(\phi\rightarrow f_0(980)\gamma)=(4.40\pm 0.21)\times 10^{-4}$
and $BR(\phi\rightarrow a_0(980)\gamma)=(7.6\pm 0.6)\times 10^{-5}$
in the Particle Physics Data Tables \cite{R12}. Achasov \cite{R15}
showed in detail that the existing data on radiative $\phi$ meson
decays support the charged-kaon loop mechanism for
$BR(\phi\rightarrow f_0(980)\gamma)$ and $BR(\phi\rightarrow
a_0(980)\gamma)$ radiative decay reactions. From the above branching
ratios within the framework of charged-kaon loop mechanism for these
decays we determine the coupling constants $g_{SK^{+}K^{-}}$ as
$g_{f_0K^{+}K^{-}}=(5.14\pm 0.12)$ GeV and
$g_{a_0K^{+}K^{-}}=(2.26\pm 0.08)$ GeV. Using these values for the
coupling constants and the values of the relevant masses taken from
the Particle Physics Data Tables \cite{R12}, if we include the
contribution of $f_0$ resonance only in the decay mechanism of the
radiative decay $\phi\rightarrow K^0\overline{K}^0\gamma$, we obtain
the result for the branching ratio as $BR(\phi\rightarrow
f_0\gamma\rightarrow K^0\overline{K}^0\gamma)=2.25\times 10^{-7}$.
On the other hand, if the contribution of $a_0$ resonance is
considered only the branching ratio is $BR(\phi\rightarrow
a_0\gamma\rightarrow K^0\overline{K}^0\gamma)=2.64\times 10^{-8}$.
Since both $f_0$ and $a_0$ resonances make a contribution to the
decay $\phi\rightarrow K^0\overline{K}^0\gamma$, when considering
their contribution to the decay rate we have to note that the
amplitudes involving $f_0$ and $a_0$ resonances interfere
destructively due to isotopic spin invariance as reflected in the
relations between the coupling constants
$g_{f_0K^{+}K^{-}}=g_{f_0K^{0}\overline{K}^{0}}$ and
$g_{a_0K^{+}K^{-}}=-g_{a_0K^{0}\overline{K}^{0}}$. Thus, if we
consider that the interference between the contributions of $f_0$
and $a_0$ resonances is destructive, we obtain for the branching
ratio the value $BR(\phi\rightarrow (a_0+f_0)\gamma\rightarrow
K^0\overline{K}^0\gamma)=9.85\times 10^{-8}$, which is somewhat
higher than the previous estimations. However, it is small enough,
we therefore conclude that this reaction will not provide a
significant background to the measurements of $\phi\rightarrow
K^0\overline{K}^0$ decay for testing CP violation.

In Fig. 2, we plot the distribution $dBR/dM_{KK}$ for the radiative
decay  $\phi\rightarrow K^0\overline{K}^0\gamma$ in the
phenomenological approach that we adopted, where we also indicate
the contributions coming from  $f_0$ resonance, $a_0$ resonances,
and the contribution resulting from the destructive interference of
these mechanisms for the coupling constants used above. Furthermore,
in Fig. 3, we show the photon spectrum $d\Gamma/dE_\gamma$ for the
process $\phi\rightarrow K^0\overline{K}^0\gamma$ for the same
amplitudes as in Fig. 2.

We note, however, that our result for the branching ratio
$B(\phi\rightarrow K^0\overline{K}^0\gamma)$ depends sensitively on
the values of the coupling constants $g_{SK^0\overline{K}^0}$ and
the masses $M_S$ of the scalar resonances $f_0$ and $a_0$ . We,
therefore, repeat our calculation using the values of the coupling
constants $g_{SK^+K^-}$ and the masses $M_S$ determined by different
experimental groups by performing fits to their data of the
radiative $\phi\rightarrow \pi^0\pi^0\gamma$ and $\phi\rightarrow
\pi^0\eta\gamma$ decays. SND Collaboration \cite{R16,R17} obtained
the values $M_{a_0}=(985.51\pm 0.8)$ MeV,
$g^2_{a_0K^{+}K^{-}}/4\pi=(0.6\pm 0.015)~~ GeV^2$, $M_{f_0}=(996\pm
1.3)$ MeV, $g^2_{f_0K^{+}K^{-}}/4\pi=(1.29\pm 0.017)~~ GeV^2$ from
their fit to SND data considering $f_0$ and $\sigma$ mixing. If we
use these values, we obtain the results $BR(\phi\rightarrow
f_0\gamma\rightarrow K^0\overline{K}^0\gamma)=1.79\times 10^{-7}$
and $BR(\phi\rightarrow a_0\gamma\rightarrow
K^0\overline{K}^0\gamma)=5.02\times 10^{-8}$ and $BR(\phi\rightarrow
(f_0+a_0)\gamma\rightarrow K^0\overline{K}^0\gamma)=4.39\times
10^{-8}$.  They also performed a fit to their data without $f_0$ and
$\sigma$ mixing in which case they obtained
$M_{a_0}=994\pm^{33}_{8}$ MeV, $g^2_{a_0K^{+}K^{-}}/4\pi=1.05\pm
^{0.36}_{0.25} ~~ GeV^2$, $M_{f_0}=(969.8\pm 4.5)$ MeV and
$g^2_{f_0K^{+}K^{-}}/4\pi=2.47\pm ^{0.73}_{0.51}~~ GeV^2$. Using
these values in our calculation give the branching ratios
$BR(\phi\rightarrow f_0\gamma\rightarrow
K^0\overline{K}^0\gamma)=2.60\times 10^{-7}$, $BR(\phi\rightarrow
a_0\gamma\rightarrow K^0\overline{K}^0\gamma)=1.03\times 10^{-7}$
and $BR(\phi\rightarrow (f_0+a_0)\gamma\rightarrow
K^0\overline{K}^0\gamma)=4.47\times 10^{-8}$. Achasov et al.
\cite{R18} also calculated the branching ratio for the decay
$\phi\rightarrow K^0\overline{K}^0\gamma$ using the above sets of
parameters describing the SND data \cite{R16,R17}, and for the first
set they obtained $BR(\phi\rightarrow (f_0+a_0)\gamma\rightarrow
K^0\overline{K}^0\gamma)=4.36\times 10^{-8}$ while for the second
set their result was $BR(\phi\rightarrow (f_0+a_0)\gamma\rightarrow
K^0\overline{K}^0\gamma)=1.29\times 10^{-8}$.

KLOE collaboration at the DA$\Phi$NE collider also studied the
decays $\phi\rightarrow \pi^0\pi^0\gamma$ and $\phi\rightarrow
\pi^0\eta\gamma$ \cite{R19,R20}. They performed two different fits
in order to measure the parameters of the scalar states. In the
first fit, they included the contribution of a possible broad scalar
$\sigma$ state as well as the intermediate $f_0(980)$ state
interfering destructively in the analysis of the data of the
$\phi\rightarrow \pi^0\pi^0\gamma$ reaction. In the second fit only
the contribution of the intermediate $f_0(980)$ state was
considered. In the first case they obtained $M_{f_0}=(973\pm 1)$
MeV, $g^2_{f_0K^{+}K^{-}}/4\pi=(2.79\pm 0.12)~~ GeV^2$ and in the
second case $M_{f_0}=(962\pm 4)$ MeV,
$g^2_{f_0K^{+}K^{-}}/4\pi=(1.29\pm 0.14)~~ GeV^2$. They also
reported the values $M_{a_0}=984.8$ MeV,
$g^2_{a_0K^{+}K^{-}}/4\pi=(0.40\pm 0.04)~~ GeV^2$. If we use the
values of  the coupling constants $g_{SK^0\overline{K}^0}$ and the
masses $M_S$ that the KLOE Collaboration obtained in the first fit,
we obtain the results $BR(\phi\rightarrow f_0\gamma\rightarrow
K^0\overline{K}^0\gamma)=2.59\times 10^{-7}$ and $BR(\phi\rightarrow
a_0\gamma\rightarrow K^0\overline{K}^0\gamma)=2.56\times 10^{-8}$
and $BR(\phi\rightarrow (f_0+a_0)\gamma\rightarrow
K^0\overline{K}^0\gamma)=1.35\times 10^{-7}$. On the other hand,
using the values they obtained in their second fit without
considering the contribution of intermediate $\sigma$ state  in our
calculation results in the branching ratios $BR(\phi\rightarrow
f_0\gamma\rightarrow K^0\overline{K}^0\gamma)=1.32\times 10^{-7}$
and $BR(\phi\rightarrow a_0\gamma\rightarrow
K^0\overline{K}^0\gamma)=2.56\times 10^{-8}$ and $BR(\phi\rightarrow
(f_0+a_0)\gamma\rightarrow K^0\overline{K}^0\gamma)=4.50\times
10^{-8}$. However, it was pointed out by Achasov et al.
\cite{R21,R22} that the KLOE data also allow
$g^2_{a_0K^{+}K^{-}}/4\pi=(0.82^{+0.81}_{-0.27})~~ GeV^2$ \cite{R21}
and $g^2_{f_0K^{+}K^{-}}/4\pi=0.62~~ GeV^2$ \cite{R22}. If we employ
these values of the coupling constants with the values of the masses
$M_{a_0}=(1003^{+32}_{-13})~~ MeV$ , $M_{f_0}=984.2~~ MeV$ and in
these analyses \cite{R21,R22} in our calculation we obtained the
results $BR(\phi\rightarrow f_0\gamma\rightarrow
K^0\overline{K}^0\gamma)=7.69\times 10^{-8}$, $BR(\phi\rightarrow
a_0\gamma\rightarrow K^0\overline{K}^0\gamma)=7.25\times 10^{-8}$,
and $BR(\phi\rightarrow (f_0+a_0)\gamma\rightarrow
K^0\overline{K}^0\gamma)=1.25\times 10^{-8}$.

We can, therefore, conclude that within the formalism we consider in
this work the branching ratio $BR(\phi\rightarrow
K^0\overline{K}^0\gamma)$ of the radiative decay $\phi\rightarrow
K^0\overline{K}^0\gamma$ is small enough so that this decay will not
cause any serious background problem for the studies of CP violation
in the $\phi\rightarrow K^0\overline{K}^0$ decay. Other effects such
as structure \cite{R23,R24} and the finite widths of scalars
\cite{R24} on the radiative $\phi$ decays have also been
investigated. Oller \cite{R24} showed that the inclusion of these
contribution does not change the conclusions of the charged-kaon
loop model.

 \begin{figure}
\epsfig{figure=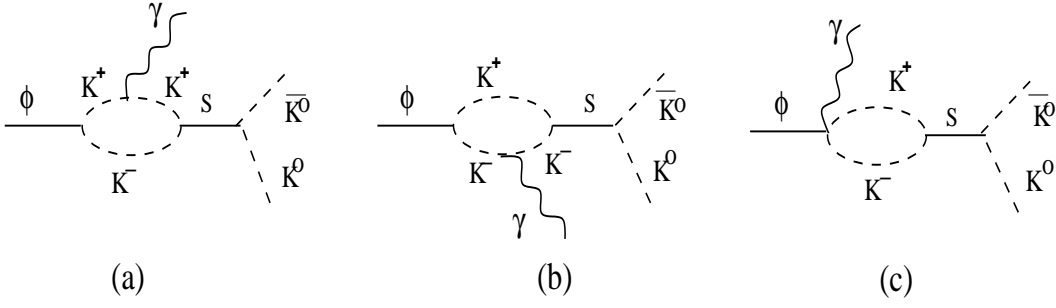,width=14cm,height=4cm} \vspace*{0.5cm}
 \caption{ Diagrams for the decay $\phi\rightarrow K^0\overline{K}^0\gamma$ where S denotes the
scalar meson resonance $f_{0}$ or $a_{0}$}
\end{figure}

\begin{figure}\hspace{1.0cm}
\epsfig{figure=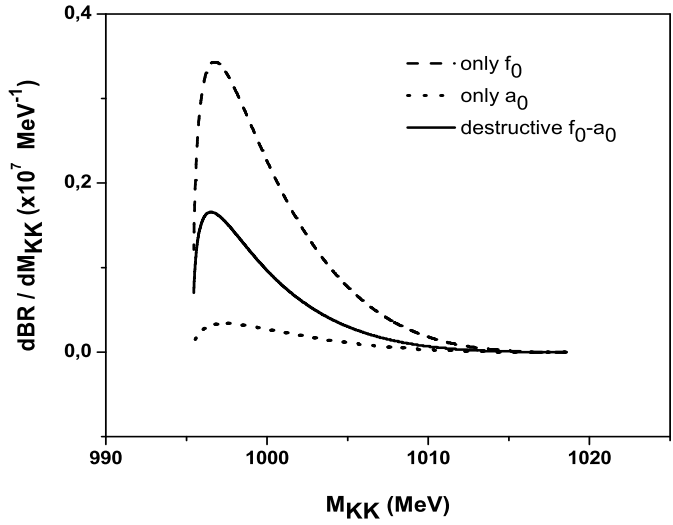,width=10cm,height=8cm} \vspace*{0.5cm}
\caption{ The distribution $dBR/dM_{KK}$ for the radiative decay
$\phi\rightarrow K^0\overline{K}^0\gamma$}
\end{figure}

\begin{figure}\hspace{1.0cm}
\epsfig{figure=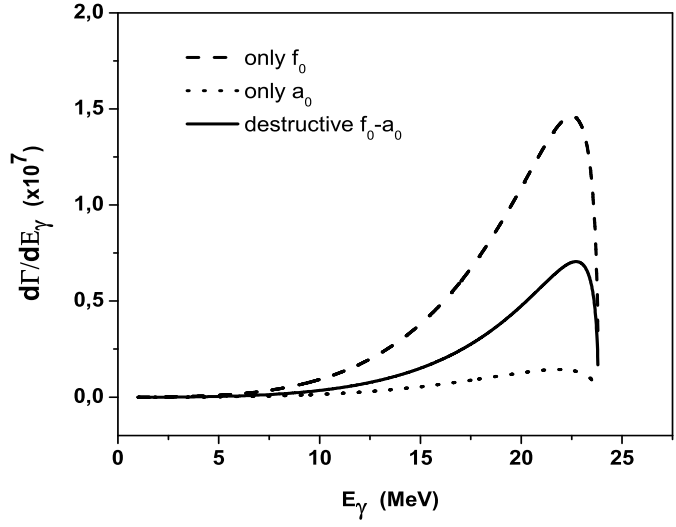,width=10cm,height=8cm} \vspace*{0.5cm}
\caption{The photon spectrum $d\Gamma/dE_\gamma$ for the process
$\phi\rightarrow K^0\overline{K}^0\gamma$  }
\end{figure}

\end{document}